\documentclass[pra,aps,amsfonts]{revtex4}

\usepackage{graphicx}
\usepackage{epsfig}

\def\qed{\leavevmode\unskip\penalty9999 \hbox{}\nobreak\hfill
     \quad\hbox{\leavevmode  \hbox to.77778em{%
               \hfil\vrule   \vbox to.675em%
               {\hrule width.6em\vfil\hrule}\vrule\hfil}}
     \par\vskip3pt}

\def\ibb #1{\leavevmode\hbox{\kern.3em\vrule
     height 1.5ex depth -.1ex width .4pt\kern-.3em\rm#1}}

\newcommand{\be}[1]{\begin{equation} #1 \end{equation}}
\newcommand{\bea}[1]{\begin{eqnarray} #1 \end{eqnarray} }
\newcommand{\ba}[2]{\left(\begin{array}{#1}#2\end{array}\right)}

\newtheorem{theorem}{Theorem}

\newcommand{\C}{{\Bbb C}}

\begin{document}

\title{On Quantum Channels.}

\author{Frank Verstraete$^{ab}$ and Henri Verschelde$^a$\\
 $^a$Department of Mathematical Physics and Astronomy, Ghent University, Belgium\\
 $^b$Department of Electrical Engineering (SISTA), KULeuven,  Belgium}

\date{February 21, 2002}

\begin{abstract}
One of the most challenging open problems in quantum information
theory is to clarify and quantify how entanglement behaves when
part of an entangled state is sent through a quantum channel. Of
central importance in the description of a quantum channel or
completely positive map (CP-map) is the dual state associated to
it. The present paper is a collection of well-known, less known
and new results on quantum channels, presented in a unified way.
We will show how this dual state induces nice characterizations of
the extremal maps of the convex set of CP-maps, and how normal
forms for states defined on a Hilbert space with a tensor product
structure lead to interesting parameterizations of quantum
channels.
\end{abstract}

\maketitle

The existence of non-local correlations or entanglement in
multipartite quantum systems \cite{EPR35,Sch35} is one of the
cornerstones on which the newly established field of quantum
information theory is build. The main gain of quantum over
classical information processing stems from the fact that we are
allowed to perform operations on entangled states: through the
quantum correlations, an operation on a part of the system affects
the whole system. One of the most challenging open problems is to
clarify and quantify how entanglement behaves when part of an
entangled state is sent through a quantum channel.

Of central importance in the description of a quantum channel or
completely positive map (CP-map) is the dual state associated to
it. This state is defined over the tensor product of the Hilbert
space itself (the input of the channel) with another one of the
same dimension (the output of the channel). It is clear that there
appears a natural tensor product structure, and indeed the notion
of entanglement will be crucial in the description of quantum
channels.

In a typical quantum information setting, Alice wants to send one
qubit (eventually entangled with other qubits)  to Bob through a
quantum channel. The channel acts linearly on the input state, and
the consistency of quantum mechanics dictates that this map be
completely positive (CP) \cite{Kra83}. This implies that the map
is of the form \cite{Cho75}
\[\Phi(\rho)=\sum_i A_i\rho A_i^\dagger.\]
Moreover the map is trace-preserving if no loss of the particle
can occur. A natural way of describing the class of CP-maps is by
using the duality between maps and states, first observed by
Jamiolkowski \cite{Jam72} and since then rediscovered by many. We
review some nice properties of CP-maps based on this dual
description, and show how to obtain the extreme points of the
convex set of trace-preserving CP-maps.

The dual state is defined on a Hilbert space that is the tensor
product of two times the original Hilbert space on which the map
acts, and is therefore naturally endowed with a notion of
entanglement. Unitary evolution for example corresponds to maximal
correlations between the in- and output state, and this kind of
evolution leads to a dual state that is maximally entangled. We
will show how normal forms derived for entangled states lead to
interesting parameterizations of CP-maps, and will discuss some
issues concerning the use of quantum channels to distribute
entanglement.

It thus turns out that the techniques developed for describing
entanglement can directly be applied for describing the evolution
of a quantum system. Concepts as quantum steering and
teleportation have a direct counterpart. A quantum channel for
example will be useful for distributing entanglement if and only
if the dual state associated to it is entangled, and optimal
decompositions of states as derived in the case of entanglement of
formation will yield very appealing parameterizations of quantum
channels.

\section{Characterization of CP-maps}

The most general evolution of a quantum system is described by a
linear CP-map \cite{Kra83}. In this section we will give a
self-contained description of CP-maps or quantum channels. Most of
the mathematics presented originate  from the seminal papers of de
Pillis \cite{deP67} and Choi \cite{Cho75}. The fact that the
evolution of quantum systems is described by linear completely
positive maps is a consequence of the assumption of the linearity
of the evolution (the complete positivity follows from consistency
arguments once the linearity is accepted).

Let us now recall some notations and useful tricks. Consider a
pure state $|\chi\rangle$ in a Hilbert space that is a tensor
product of two Hilbert spaces of dimension $n$
\[|A\rangle=\sum_{ij}^n a_{ij}|i\rangle|j\rangle.\] Define
\[|I\rangle=\sum_i^n|i\rangle|i\rangle\] an unnormalized maximally
entangled state and $A$ the operator with elements $\langle
i|A|j\rangle=a_{ij}$, then
\[|A\rangle=A\otimes I_n|I\rangle.\]
Moreover it holds that
\[X\otimes Y|A\rangle=XA\otimes Y|I\rangle=XAY^T\otimes
I_n|I\rangle=I_n\otimes YA^TX^T|I\rangle.\]

The symbol $|I\rangle$ will solely be used to denote the
unnormalized maximally entangled state $|I\rangle=\sum_i
|ii\rangle$. We are now ready for the following fundamental
Theorem of de Pillis\cite{deP67}:
\begin{theorem}
A linear map $\Phi$ acting on a matrix $X$ is Hermitian-preserving
if and only if there exist operators $\{A_i\}$ and real numbers
$\lambda_i$ such that
\[\Phi(X)=\sum_i\lambda_i A_iXA_i^\dagger\]
\end{theorem}
{\em Proof:} Suppose the map $\Phi$  acts on a $n\times n$ matrix.
Then due to linearity, $\Phi$ is completely characterized if we
know how it acts on a complete basis of $n\times n$ matrices, for
example on all matrices $|e_i\rangle\langle e_j|$, $1\leq i,j\leq
n$ with ${|e_i\rangle}$ a complete orthonormal base in Hilbert
space. Let us define the $n^2\times n^2$ positive matrix
\be{|I\rangle\langle I|=\ba{ccc}{|e_1\rangle\langle
e_1|&\cdots&|e_1\rangle\langle
e_n|\\\cdots&\cdots&\cdots\\|e_n\rangle\langle
e_1|&\cdots&|e_n\rangle\langle e_n|},} being the matrix notation
of a maximally entangled state in a $n\otimes n$ Hilbert space. It
follows that all the information of a map $\Phi$ is encoded in the
state \be{\rho_\Phi=I_n\otimes\Phi(|I\rangle\langle I|),} as the
$n^2$ $n\times n$ blocks represent exactly the action of the map
on the complete basis $|e_i\rangle\langle e_j|$. If $\Phi$ is
Hermitian-preserving, then $\Phi(|e_i\rangle\langle e_j|)$ has to
be equal to the Hermitian conjugate of $\Phi(|e_j\rangle\langle
e_i|)$, and this implies that $\rho_\Phi$ is Hermitian. Let us
therefore consider the eigenvalue decomposition of
$\rho_\Phi=\sum_i\lambda_i |\chi_i\rangle\langle\chi_i|$. Using
the trick $|A\rangle=(A\otimes I)|I\rangle$, we easily arrive at
the conclusion that $\Phi(X)=\sum_i\lambda_i A_iXA_i^\dagger$,
where $\{\lambda_i\}$ are the eigenvalues and where the operators
$\{A_i^T\}$ are the reshaped versions of the eigenvectors of
$\rho_\Phi$.\qed

A central ingredient in the proof was the introduction of the
matrix \[\rho_\Phi=I_n\otimes\Phi(|I\rangle\langle I|)\]  with
$|I\rangle=\sum_i|i\rangle|i\rangle$  a maximally entangled state.
We define this Hermitian matrix  $\rho_\Phi$ as being the dual
state corresponding to the map $\Phi$. It was already explained
that it encodes all the information about the map, and its
eigenvectors give rise to the operators $A_i$. The above lemma
characterizes all possible Hermitian preserving maps, and
therefore surely all positive and completely positive maps. For
example, let us consider the positive map that corresponds to
taking the transpose of the density operators of a qubit:
\bea{\lambda_1&=&1\hspace{1cm}A_1=\ba{cc}{1&0\\0&0}\\
\lambda_2&=&1\hspace{1cm}A_2=\ba{cc}{0&0\\0&1}\\
\lambda_3&=&1\hspace{1cm}A_3=\ba{cc}{0&1\\1&0}/\sqrt{2}\\
\lambda_4&=&-1\hspace{1cm}A_4=\ba{cc}{0&1\\-1&0}/\sqrt{2}}

Not all Hermitian-preserving maps are physical in quantum
mechanics however: if a map acts on a subsystem, then it should
conserve positivity of the complete density operator. This extra
assumption leads to the condition of complete positivity, meaning
that $I_m\otimes \Phi$ is positive for all $m$. Of course, this
implies that the dual state $\rho_\Phi$ is not only Hermitian but
also positive (i.e. all its eigenvalues are positive), as it is
defined as the action of the map $I_n\otimes\Phi$ on a maximally
entangled state.  The positive eigenvalues can then be absorbed
into the (Kraus) operators $\{A_i\}$, and we have therefore proven
the Kraus representation Theorem (Choi\cite{Cho75}):
\begin{theorem}
A linear map $\Phi$ acting on a density operator $\rho$ is
completely positive  if and only if there exist operators
$\{A_i\}$  such that
\[\Phi(\rho)=\sum_i A_i\rho A_i^\dagger.\]
\end{theorem}
Remarks: \begin{itemize} \item  A CP-map is trace-preserving iff
$\sum_iA_i^\dagger A_i=I_n$; this property is easily verified
using the cyclicity of the trace. In terms of the (unique) dual
state $\rho_\Phi$ associated to the map $\Phi$, this
trace-preserving condition amounts to:
\[Tr_{2}(\rho_\Phi)=I_n.\] Here the notation $Tr_2$ means the
partial trace over the second subsystem. A CP-map is furthermore
called bistochastic if also the condition
\[Tr_{1}(\rho_\Phi)=I_n\] holds; this property is equivalent to the
fact that the map is identity-preserving, i.e. $\Phi(I_n)=I_n$.

\item The dual state $\rho_\Phi$ corresponding to a CP-map $\Phi$
is uniquely defined. The Kraus operators are obtained by
considering the columns of a square root of $\rho_\Phi$ ($A_i$ is
obtained by making a matrix out of the i'th column of a square
root of $X$, with $\rho_\Phi=XX^\dagger$). As the square root of a
matrix is not uniquely defined, the Kraus operators are not
unique. Each different ``square root" $X$ of $\rho_\Phi$
($\rho_\Phi=XX^\dagger$) gives rise to a different set of
equivalent Kraus operators. This implies that all equivalent sets
of Kraus operators are related by an isometry, and that the
minimal number of Kraus operators is given by the rank of the
density operator $\rho_\Phi$. Therefore we define the rank of a
map to be the rank of the dual operator $\rho_\Phi$. This rank is
bounded above by $n^2$ with $n$ the dimension of the Hilbert
space. A unique Kraus representation  can be obtained by for
example enforcing the Kraus operators to be orthogonal, as these
would correspond to the unique eigenvectors of $\rho_\Phi$. Note
that a similar reasoning applies to all Hermitian preserving and
all positive maps, although there an additional sign should be
taken into account.

\item By construction, we have proven that a map $\Phi$ acting on
a n-dimensional Hilbert space is completely positive iff
$I_n\otimes \Phi$ is positive: there is no need to consider
auxiliary Hilbert spaces with dimension larger than the original
one. The reasoning is as follows: if $I_n\otimes  \Phi$ is
positive, then $\rho_\Phi$ is positive, and therefore $\Phi$ has a
Kraus representation, which implies complete positivity.

\item Suppose $\Phi$ is positive but not completely positive. Then
there exists a completely positive map $\tilde{\Phi}$ and a
positive scalar $\epsilon$ such that
\[\Phi(\rho)=(1+n\epsilon)\tilde{\Phi}(\rho)-\epsilon{\rm Tr}(\rho)I_n.\]
The proof of this fact is elementary: take $\epsilon$ to be the
opposite of the smallest eigenvalue of $\rho_\Phi$ (this
eigenvalue is negative as otherwise $\Phi$ would be completely
positive), and define the CP-map
$\tilde{\Phi}(\rho)=(\Phi(\rho)+n\epsilon{\rm
Tr}(\rho)I/n)/(1+n\epsilon)$ (this map is completely positive
because the dual state $A_{\tilde{\Phi}}$ associated to it is
positive and has therefore a Kraus representation). Note that the
whole reasoning is also valid for general Hermitian-preserving
maps. As an example, consider again the transpose map on  a qubit.
Then it can be checked that the minimal value of $\epsilon$ is $1$
(this is true for the PT operation in arbitrary dimensions) and
that the Kraus operators corresponding to $\tilde{\Phi}$ become
\[\{A_i\}=\{\sqrt{\frac{2}{3}}\ba{cc}{1&0\\0&0},\sqrt{\frac{2}{3}}\ba{cc}{0&0\\0&1},\sqrt{\frac{1}{3}}\ba{cc}{0&1\\1&0}/\sqrt{2}\}.\]

\item To make the duality between maps and states more explicit,
it is useful to consider the following identity:
\be{\Phi(\rho)=Tr_2\left(\rho_\Phi^{T_1}(\rho\otimes
I_n)\right),\label{dual}} where $T_1$ means partial transposition
with relation to the first subsystem. This can be proven by
explicitly writing the map $\Phi$ into Kraus operator form, and
exploiting the cyclicity of the trace. Due to the partial
transpose condition of Peres \cite{Per96}, it is clear that
$\rho_\Phi^{T_1}$ will typically not longer be positive. This
identity is very useful, and was used in  the section on optimal
teleportation with mixed states.
\end{itemize}

\section{Extreme points of CP-maps}

The set of completely positive maps is a convex set: indeed, if
$\Phi_1$ and $\Phi_2$ are CP-maps, then so is
$x\Phi_1+(1-x)\Phi_2$. Due to the one to one correspondence
between maps $\Phi$ and states $\rho_\Phi$, it is trivial to
obtain the extreme points of the set of completely positive maps:
these are the maps with one Kraus operator, corresponding to
$\rho_\Phi$ having rank 1.

If however we consider the convex set of trace-preserving maps,
the characterization of extreme points becomes more complicated.
The knowledge of the set of extreme points of the trace-preserving
CP-maps  is very interesting from a physical perspective in the
following way: suppose one has a multipartite state of qudits and
one wants to maximize some convex functional of the state (e.g.
the fidelity, ...) by performing local operations. Due to
convexity, the optimal operation will correspond to an extreme
point of the set of trace-preserving maps.

Let us now characterize all extremal trace-preserving maps:
\begin{theorem}\label{thep}
Consider a TPCP-map $\Phi$ acting on a Hilbert space of dimension
$n$ and of rank $m$. Consider the dual state
$\rho_\Phi=XX^\dagger$ with $X$ a $n^2\times m$ matrix, and the
$n^2$ matrices $X_i=X^\dagger (\sigma_i\otimes I_n)X$ (the
matrices $\{\sigma_i\}$ form a complete basis for the Hermitian
$n\times n$ matrices). Then $\Phi$ is extremal if and only if
$m\leq n$ and if the set of linear equations $\forall i: {\rm
Tr}(QX_i)=0$ has only the trivial solution $Q=0$.

This condition is equivalent to the following one given by
Choi\cite{Cho75}: given $m^2$ Kraus operators $\{A_i\}$ of a map
$\Phi$, then the map is extremal iff the $m^2$ matrices
$\{A_i^\dagger A_j\}$, $1\leq i,j\leq m$ are linearly independent.
\end{theorem}
{\em Proof:} The map $\Phi$ is extremal if and only if there does
not exist a $R$ with the property that $RR^\dagger\neq I$ and such
that ${\rm Tr}_2(XRR^\dagger X^\dagger)=I$. This condition is
equivalent to the fact that the set of equations
\[{\rm Tr}\left(X\underbrace{(RR^\dagger-I)}_{Q}X^\dagger \sigma_i\otimes
I\right)=0\] does only have the trivial solution $Q=0$. As there
are $n^2$ independent generators $\sigma_i$ and due to the fact
that $Q$ has $m^2$ degrees of freedom, it is immediately clear
that there will always be a non-trivial solution if $m>n$, ending
the proof.

It remains to be proven that he condition obtained is equivalent
to the one derived\footnote{Actually, Choi derived the different
problem of characterizing the extremal points of the (not
necessarily trace-preserving) CP-maps that leave the identity
unaffected, but his arguments are readily translated to the
present situation. Note also that his proof was much  more
involved.} by Choi \cite{Cho75}. This can be seen as follows: the
condition ${\rm Tr}_2(XRR^\dagger X^\dagger)=I$ is equivalent to
the condition $\sum_{jk}A_k^\dagger A_j(\sum_i
R_{ji}R^*_{ki}-\delta_{jk})=0$ (this is readily obtained using the
trick $|A\rangle=A\otimes I|I\rangle$). Therefore a nontrivial
solution of $Q$ is possible iff the set of matrices $\{A_i^\dagger
A_j\}$, $1\leq i,j\leq m$ are linearly dependent.\qed

Note that the given proof is constructive and can therefore be
used for decomposing a given TPCP-map into a convex combination of
extremal maps: once a non-trivial $Q$ and therefore $R$ is
obtained, one can scale it such that $RR^\dagger\leq I$, and
define another $S=\sqrt{I-RR^\dagger}$. This $S$ is guaranteed to
be another trace-preserving map up to a constant factor, and the
original map is the sum of the maps parameterized by $XRR^\dagger
X^\dagger$ and $XSS^\dagger X^\dagger$.

All TPCP maps $\Phi$ of rank $1$ are of course extreme and
correspond to unitary dynamics. One easily verifies that this
implies that the dual $\rho_\Phi$ is a maximally entangled state.
The intuition behind this is as follows: by equation (\ref{dual}),
$\rho_\Phi$ characterizes the correlation between the output and
the input of the channel. Maximal correlation happens iff the
evolution occurs  reversibly and thus unitarily, and therefore
corresponds to maximal ``entanglement" between in- and output. We
will explore this connection between maps and entanglement more
thoroughly in the following section.

One could go one step further, and try to characterize all extreme
points of the convex set defined by all trace-preserving channels
for which the extra condition holds that $\Phi(\rho_1)=\rho_2$
with $\rho_1$ and $\rho_2$ given density operators. (Note that
$\rho_1$ and $\rho_2$ can be chosen completely arbitrary, as there
will always exist at least one TPCP-map that transforms a given
state into another given one: consider for example the map with
its associated dual state $\rho_\Phi=I\otimes\rho_2$.)
Bistochastic channels are a special subset of this convex set of
maps (in that case $\rho_1=\rho_2\simeq I$). An adaption of
Theorem \ref{thep} leads to the following:

\begin{theorem}
Consider the convex set of trace-preserving CP-maps $\Phi$ for
which $\Phi(\rho_1)=\rho_2$ with $\rho_1,\rho_2$ given. Suppose
$\Phi$ is of rank $m$, its dual state is $\rho_{\Phi}=XX^\dagger$
with $X$ a $n\times n$ matrix,  and that there are  $m$ Kraus
operators $\{A_i\}$. Then this map is extremal if and only if the
set of $2m^2$ linear equations \be{{\rm
Tr}(QX^\dagger(\sigma_i\otimes I)X)=0\hspace{1cm}{\rm
Tr}(QX^\dagger(\rho_1^T\otimes \sigma_i)X)=0} has only the trivial
solution $Q=0$, or equivalently if and only if the $m^2$ operators
$\{A_i^\dagger A_j\oplus A_j\rho_1 A_i^\dagger\}$ ($1\leq i,j\leq
m$) are linearly independent.
\end{theorem}
Proof: The proof is completely analogous to the proof of Theorem
\ref{thep}, but here we have the extra condition
\[{\rm Tr}\left(X(RR^\dagger -I)X(\rho_1^T\otimes \sigma_i)\right)=0.\]
In terms of Kraus operators, this  additional condition becomes
\[\sum_{kj}A_j\rho_1 A_k^\dagger(\sum_i
R_{ji}R_{ki}^*-\delta_{jk})=0\] which ends the proof. \qed

A similar Theorem was stated by Landau and Streater \cite{LS93} in
the special  case of bistochastic maps. In analogy with the
conclusions of Theorem \ref{thep}, we conclude that the number of
Kraus operators in an extremal TPCP-map of the kind considered in
the above Theorem is bounded by $\lfloor\sqrt{2n^2}\rfloor$.

Let us for example consider the case of qubits. Then the rank of
an extremal $\Phi$ is bounded by $2$, and extremal rank 2
TPCP-maps obeying the condition $\Phi(\rho_1)=\rho_2$ typically
exist. There is however a notable exception if $\rho_1=\rho_2=I/2$
(i.e. when $\Phi$ is bistochastic): a bistochastic qubit map has a
corresponding dual $\rho_\Phi$ that is Bell-diagonal. A
Bell-diagonal state is a convex sum of maximally entangled states,
and therefore a rank 2 bistochastic map cannot be extremal. Note
however that this is an accident, and for Hilbert space dimensions
larger than 2 there exist extremal bistochastic channels that are
not unitary \cite{LS93}. Sometimes the name ``unital" is also used
instead of ``bistochastic". The foregoing argument however shows
that this terminology is not completely justified.

One could now add more constraints $\Phi(\rho_{2i})=\rho_{2i+1}$,
and this would lead to similar conditions for extremality in terms
of the Kraus operators. Note however that the $\rho_i$ appearing
in the constraints cannot be chosen completely arbitrary, as in
general non-compatible constraints can arise due to the complete
positivity condition on the physical maps (Deciding whether a set
of conditions  $\Phi(\rho_{2i})=\rho_{2i+1}$ is physical can be
solved using the techniques of semidefinite programming
\cite{VB96}).

Let us now formulate  another interesting Theorem:
\begin{theorem}\label{thepp}
Given a Hilbert space of dimension $n$ and a trace-preserving map
$\Phi$ of rank $m\leq n$, then there exist pure states
$|\psi\rangle$ such that $\Phi(|\psi\rangle\langle\psi|)$ are
states of rank $m-1$.
\end{theorem}
Proof: Let us first consider the case $m=n$, and define $m$ Kraus
operators $\{A_i\}$ corresponding to $\Phi$. Given a pure state
$|\psi\rangle$, then $\Phi$ maps this state to one that is not
full rank iff there exists a pure state $|\chi\rangle$ such that
\[\langle\chi|\Phi(|\psi\rangle\langle\psi|)|\chi\rangle=0=\sum_i|\langle\chi|A_i|\psi\rangle|^2.\]
Writing $|\chi\rangle =\sum_i y_i|i\rangle$, $|\psi\rangle=\sum_i
x_i|i\rangle$ and $\langle j|A_i|k\rangle=A^j_{ik}$, then the
previous equation amounts to solving the following set of bilinear
equations:
\[\forall i=1:n,  \sum_{k=1}^n (\sum_{j=1}^{m=n} x_jA^j_{ik})y_k=0.\]
This set of equations always has a non-trivial solution. Indeed,
the parameters $x_j$ can always be chosen such that the matrix
$\tilde{A}=\sum_j x_jA^j_{ik}$ is singular (if all $A_i$ are full
rank then this can be done by fixing all but one of them, and then
choosing the remaining parameter such that the determinant
vanishes; if one of the $A_i$ is rank deficient then the solution
is of course direct). Then the parameters $y_k$ can be chosen such
that the vector $y$ is in the right kernel of $\tilde{A}$ (the
right kernel is not zero-dimensional as the dimension of the
matrix $\tilde{A}$ is $n\times n$), and therefore
$\Phi(|\psi\rangle\langle\psi|)$ is not full rank. If $m<n$, then
the right kernel of $\tilde{A}$ is at least $n-m+1$ dimensional,
such that $n-m+1$ linearly independent $|\chi\rangle$ can be found
such that
$\langle\chi|\Phi(|\psi\rangle\langle\psi|)|\chi\rangle=0$, which
ends the proof.\qed

In general , it is thus proven that one can always find states
$|\psi\rangle$ such that the rank of
$\Phi(|\psi\rangle\langle\psi|)$ is smaller than the rank of the
map, which is surprising. Note that the bound in the Theorem is
generically tight, i.e. the minimal rank of the output state will
typically be $m-1$; this follows from the fact that decreasing the
rank of the  matrix $\tilde{A}$ with two units would need
$n(n-1)/2$ independent degrees of freedom, while there are only
$n-1$ available.

Note that extremal TPCP-maps always fulfil the conditions of the
Theorem. In particular, extremal qubit channels are generically of
rank 2, and the previous Theorem implies that there always exist
pure states that remain pure after the action of a rank 2 extremal
map (This was also observed by Ruskai et al.\cite{RSW02}).

The above Theorem has also some consequences for the study of
entanglement.  Applying the foregoing proof to the dual state
$\rho_\Phi$, we can easily prove the following: if the rank of a
mixed state $\rho$ defined in a $n\times n$ dimensional Hilbert
space is given by $m\leq n$, then there always exist at least
$(n-m+1)$ linearly independent product states orthogonal to it.

Let us now consider an example of the use of extremal maps.
Suppose we want to characterize the optimal local trace-preserving
operations that one has to apply locally to each of the qubits of
a 2-qubit entangled mixed state, such as to maximize the fidelity
(i.e. the overlap with a maximally entangled state). This problem
is of  interest in the context of teleportation \cite{BBC93,HHH99}
as the fidelity of the state used to teleport is the standard
measure of the quality of teleportation. Badziag and the
Horodecki's \cite{BHH00} discovered the intriguing property that
the fidelity of a mixed state can be enhanced by applying an
amplitude damping channel to one of the qubits. This is due to the
fact that the fidelity is both dependent on the quantum
correlations and on the classical correlations, and enhancing the
classical correlations by mixing (and hence losing quantum
correlations) can sometimes lead to a higher fidelity.

With the help of the previous analysis of extremal maps, we are in
the right position to find the optimal trace-preserving map that
maximizes the fidelity.  Indeed, the optimization problem is to
find the trace-preserving CP-maps $\Phi_A,\Phi_B$ such as to
maximize the fidelity $F$ defined as
\begin{equation}
F(\rho,\Phi_A,\Phi_B)=\langle\psi|\Phi_A\otimes
\Phi_B(\rho)|\psi\rangle={\rm
Tr}\left\{\rho\left(\Phi_A^\dagger\otimes\Phi_B^\dagger(|\psi\rangle\langle\psi|)\right)\right\}
\end{equation} with $|\psi\rangle$ the maximally
entangled state. This problem is readily seen to be jointly convex
in $\Phi_A$ and $\Phi_B$, and therefore the optimal strategy will
certainly consist of applying extremal (rank 2) maps
$\Phi_A,\Phi_B$. As we just have derived an easy parametrization
of these maps, it is easy to devise a numerical algorithm that
will yield the optimal solution.

Note that the problem, although convex in $\Phi_A$ and $\Phi_B$,
is bilinear and therefore can have multiple (local) maxima. This
problem disappears when only one party (Alice or Bob) applies a
map (i.e. $\Phi_B=I$). This problem was studied in more detail by
Rehacek et al.\cite{RHF01}, where a heuristic algorithm was
proposed to find the optimal local trace-preserving map to be
applied by Bob. As the optimization problem is however convex, the
powerful techniques of semidefinite programming \cite{VB96} should
be applied, for which an efficient algorithm exists that is
assured to converge to the global optimum. Indeed, due to
linearity the problem now consists of finding the 2-qubit state
$\rho_{\Phi^\dagger}\geq 0$ with constraint ${\rm
Tr_B}(\rho_{\Phi^\dagger})=I$ such that the fidelity is maximized.
As we already know, the algorithm will converge to a  $\rho_\Phi$
of maximal rank 2 in the case of qubits. Exactly the same
reasoning holds for systems in higher dimensional Hilbert spaces:
if only one party is to apply a trace-preserving operation to
enhance the fidelity, the above semidefinite program will produce
the optimal local map that maximally enhances the fidelity.

Other situations in which extremal maps will be encountered are
for example the problem of optimal cloning\cite{BH96,Cer00,AD02}:
given an unknown input state $\rho$, one wants to construct the
optimal trace-preserving CP-map such as to yield an output for
which the fidelity with $\rho\otimes\rho$ is maximal. This can
again be rephrased as a semidefinite program whose unique solution
will be given by an extremal trace-preserving CP-map.

\section{Quantum channels and entanglement}

The physical interpretation of the dual state corresponding to a
CP-map or quantum channel is straightforward. It is the density
operator that corresponds to the state that can be made as
follows: Alice prepares a maximally entangled state $|I\rangle$,
and sends one half of it to Bob through the channel $\Phi$. This
results into $\rho_\Phi$.

A perfect quantum channel is unitary and the corresponding state
$\rho_\Phi$ is a maximally entangled state. This corresponds to
the case of perfect transmission of qudits, and indeed a maximally
entangled state is the state with perfect quantum correlations.
Consider now a completely depolarizing channel. In that case it is
possible to transmit a classical bit perfectly, and indeed
$\rho_\Phi$ corresponds to a separable state with maximal
classical correlations. As a third example, consider the complete
amplitude damping channel. Then $\rho_\Phi$ is a separable pure
state with no correlations whatever between Alice and Bob. It is
therefore clear that the study of the character of correlation
present in the quantum state $\rho_\Phi$ tells us a lot about the
character of the quantum channel.

This way of looking at quantum channels gives a nice way of
unifying statics and dynamics in one framework: the future is
entangled (or at least correlated) with the past. Just as a
measurement in the future gives us information about the prepared
system (through the use of the quantum Bayes rule), a measurement
on Bob's side enables Alice to refine her knowledge of her local
system (through the use of the quantum steering
Theorem)\footnote{In some sense one could argue that this was
expected due to the fact that space and time play analogous roles
in the theory of relativity. It is very nice however that in the
non-relativistic case considered here, the duality is already
present. This gives hope that it should be possible to generalize
the current findings to the relativistic case.}. It is therefore
clear that the description of entanglement will shed new light on
the question of describing correlations between the states of the
same system at two different instants of time, and vice-versa.
Therefore we expect that many useful results concerning
entanglement can directly be applied to quantum channels. On the
other hand, a lot of work has been done concerning the
quantification of the classical capacity of a quantum channel.
These results offer a nice starting point for the study of
classical correlations present in a quantum state.

\subsection{Quantum  capacity}

The quantum capacity of a quantum channel is related to the
asymptotic number of uses of the channel needed for obtaining
states whose fidelity tends to one. To transmit quantum
information with high fidelity, one indeed needs almost perfect
singlets. It is immediately clear that ideas of entanglement
distillation will be crucial: sending one part of an EPR through
the channel will result in a mixed state, and these mixed states
will have to be purified.

Let us first establish a result that was already intrinsically
used  by many \cite{BDS96,HHH99,Rai01,CDK01}:

\begin{theorem}\label{entbre}
A quantum channel $\Phi$ can be used to distribute entanglement if
and only if $\rho_\Phi$ is entangled. If $\rho_\Phi$ is separable,
then the Kraus operators of the map $\Phi$ can be chosen to be
projectors, and the map $\Phi$ is entanglement breaking.
\end{theorem}
Proof: The if part is obvious, as $\rho_\Phi$ is the state
obtained by sending one part of a maximally entangled state
through the channel. To prove the only if part, assume that
$\rho_\Phi$ is separable. Then all Kraus-operators can be chosen
to be projectors (corresponding to the decomposition with
separable pure states), destroying all entanglement.\qed

It is also possible to make a quantitative statement:

\begin{theorem}\label{thme}
Suppose we want to use the channel $\Phi$ to distribute
entanglement by sending one part of an entangled state through the
channel. The maximal attainable fidelity (i.e. overlap with a
maximally entangled state) corresponds to the largest eigenvalue
of $\rho_\Phi$. This maximal fidelity is obtained if Alice sends
one half of the state described by the eigenvector of $\rho_\Phi$
corresponding to its largest eigenvalue.
\end{theorem}
Proof: Suppose Alice prepares the entangled state $|\chi\rangle$
and sends the second part to Bob through the channel $\Phi$ with
Kraus-operators $\{A_i\}$. We want to find the state
$|\chi\rangle$ such that \be{\langle I|\sum_i I\otimes
A_i|\chi\rangle\langle\chi|I\otimes
A_i^\dagger|I\rangle=\langle\chi|\rho_\Phi|\chi\rangle} is
maximized, which immediately gives the stated result.\qed

The above result is amazing: it tells us that it is not always the
best strategy to send one part of a maximally entangled state
through the channel. It would be tempting to conjecture that the
entanglement of distillation of the obtained state represents the
quantum capacity of the given channel.

Note that the eigenvalues and eigenvectors of $\rho_\Phi$ got an
appealing interpretation: these represent the fidelities that are
obtained by sending one half of the eigenvectors through the
channel. Note also that the reduction criterion \cite{HH99,CAG99},
\[I\otimes\rm{Tr}_2(\rho_\Phi)-\rho_\Phi=\frac{I}{n}-\rho_\Phi\]
implies that $\rho_\Phi$ is entangled if its largest eigenvalue
exceeds $1/n$. This is of course in complete accordance with the
previous Theorem, as the maximal fidelity for a separable state is
also given by $1/n$.

A more sophisticated treatment of the quantum capacity of a
quantum channel would involve ideas of coding and of quantum error
correction, although only partial results have been obtained yet;
the following is an incomplete list of papers where interesting
results have been obtained
\cite{BDS96,Sch96,DSS98,Win99,BKN00,Ham01,Ham02}.

\subsection{Classical Capacity}

Let us now move towards the well-studied problem of classical
capacity of a quantum channel. The central result is the Holevo-
Schumacher- Westmoreland Theorem \cite{Hol98,SW97}, which tells us
that the classical product state capacity of a quantum channel
$\Phi$ is given by
\be{\chi(\Phi)=\max_{p_j,\rho_j}\left\{S(\Phi(\sum_jp_j\rho_j))-\sum_jp_jS(\Phi(\rho_j))\right\}.\label{HSW}}
Let us now ask the following question: what would be the analogy
and the interpretation of this formula in the dual picture of
states $\rho_\Phi$? Using formula (\ref{dual}), it holds that
\[\Phi(\rho_j)={\rm Tr}_1(\rho_\Phi(\rho_j^T\otimes I)).\]
Suppose Alice and Bob share the state $\rho_\Phi$. Then the above
formula describes how Bob has to update his local density operator
when Alice did a measurement with corresponding POVM-element
$\rho_j^T$. Reasoning along the lines of the HSW-Theorem, the
natural interpretation would now be  that formula (\ref{HSW}) will
give us a measure of how much (secret) classical randomness Alice
and Bob can create using the state $\rho_\Phi$: if Alice
implements a POVM measurement with elements $\{p_j,\rho_j^T\}$,
this drives the system at Bob's side into a particular direction,
and a measurement of Bob will reveal some information about the
(random) outcome of Alice. Note that we interpret the presence of
a bipartite state as being a particular kind of quantum channel.
Note that the question of creating shared randomness has also been
discussed in  \cite{THL02,CMS02}.

The foregoing discussion suggests the following definition for the
classical random correlations $C^{cl}$ present in a quantum state
$\rho$:
\bea{C^{cl}_B(\rho_{AB})&=&\max_{\{E_j\}}S(\rho_B)-\sum_j p_jS(\rho_B^j)\label{Ecl}\\
p_j&=&{\rm Tr}{\rho(E_j\otimes I)}\\
\rho_B^j&=&\frac{1}{p_j}{\rm Tr}_1\left(\rho(E_j\otimes
I)\right).} Here $\{E_j\}$ presents the elements of the POVM
implemented by Alice. Observe that there is an asymmetry in the
definition, in that $C^{cl}_A$ is not necessarily equal to
$C^{cl}_B$. This definition coincides with the one given by
Henderson and Vedral \cite{HV01}, where they introduced this
measure because it fulfilled the condition of monotonicity under
local operations.

In general, the classical mutual information obtained by the
actions of Alice and Bob to obtain classical randomness will be
smaller than the derived quantity (\ref{Ecl}), as coding is needed
to achieve the Shannon capacity. This coding could be implemented
by doing joint measurements, but we do not expect that the upper
bound is tight; a better rate could be obtained if also public
classical communication is allowed (A. Winter, unpublished).

\section{One-qubit channels}
In the case of qubit channels, much more explicit results can be
obtained, due to the fact that we have a fairly good insight into
the properties of mixed states of two qubits. In this section we
highlight some questions about qubit channels that can be solved
analytically.

Recall formula (\ref{dual})
\be{\Phi(\rho)=Tr_1\left(\rho_\Phi^{T_1}(\rho\otimes I_n)\right)}
which is almost exactly the same expression as if Alice were
measuring the POVM-element $\rho$ on the joint state $\rho_\Phi$;
the difference it that the partial transpose of this state has to
be taken. It is now natural to look at the R-picture of the dual
state $\rho_\Phi$ associated to the map \cite{VDD01b}, where
$\rho$ is parameterized by a real $4\times 4$ matrix
\[R_{ij}={\rm Tr}\left(\rho\sigma_i\otimes \sigma_j\right),\]
$0\leq \sigma_i\leq 3$. In the R-representation, a partial
transpose corresponds to a multiplication of the third column or
row with a minus sign. Let us therefore define $R_\Phi$ to be the
parameterization of $\rho_\Phi^{T_1}$ in the $R$-picture, i.e. the
R-picture of $\rho_\Phi$ in which the third row is multiplied by
$-1$. Note that the first row of $R_\Phi$ is given by $[1;0;0;0]$,
as this corresponds to the trace-preserving condition.

If $x$ is the Bloch vector corresponding, then  the action of the
map with corresponding $\rho_\Phi^{T_1}$ or $R_\Phi$ is the
following: \be{\ba{c}{1\\x'}=R_\Phi\ba{c}{1\\x}}.  One can easily
prove that the image of the Bloch sphere yields an ellipsoid,
where the local density operator of Alice is represented by the
center of the ellipsoid. This implies that the knowledge of the
ellipsoid corresponds to the complete knowledge of the quantum
channel up to local unitaries at the input. (Note that not all
ellipsoids correspond to physical maps, but that there is some
restriction on the ratio of the axis).

Let us now consider the analogue of local unitary (LU) and local
filtering (SLOCC) equivalence classes as known for mixed states of
two qubits \cite{VDD01b}. What we are looking for are normal forms
$\Omega$ (where $\Omega$ is a map) such that
$\Phi(\rho)=B\Omega(A\rho A^\dagger)B^\dagger$ with $A,B\in SU(2)$
or $\in SL(2,\C)$.

The LU case is very easy: each $R_\Phi$ can be brought into the
unique form
\[R_\Phi=\ba{cccc}{1&0&0&0\\x&\lambda_1&0&0\\y&0&\lambda_2&0\\z&0&0&\pm\lambda_3}\]
by local unitary transformations, where
$\lambda_1\geq\lambda_2\geq|\lambda_3|$ and $x,y\geq 0$; one just
has to take the singular value decomposition of the lower $3\times
3$ block of $R$, taking into account that the orthogonal matrices
have determinant $+1$ (see also Fujiwara and Algoet \cite{FA99}
and King and Ruskai \cite{KR01} for a different approach but with
the same result).

Let us next move to SLOCC equivalence classes; it is clear that
the Lorentz singular value decomposition \cite{VDD01b} is all we
need:

\begin{theorem}
Given a 1-qubit trace-preserving CP-map $\Phi$ and its dual
$R_\Phi$. Then the SLOCC normal form $\Omega$ of $R_\Phi$ is
proportional to  one of the following unique normal forms:
\be{\ba{cccc}{1&0&0&0\\0&s_1&0&0\\0&0&s_2&0\\0&0&0&s_3}\hspace{.2cm}\ba{cccc}{1&0&0&0\\0&x/\sqrt{3}&0&0\\0&0&x/\sqrt{3}&0\\2/3&0&0&1/3}
\hspace{.2cm}\ba{cccc}{1&0&0&0\\0&0&0&0\\0&0&0&0\\1&0&0&0}\nonumber.}
Here $1\geq s_1\geq s_2\geq |s_3|$, $1-s_1-s_2-s_3\geq 0$ and
$0\leq x\leq 1$. For maps with a normal from of the first kind,
one can choose the Kraus operators equal to
\be{\{A_i\}=\{p_0A\sigma_0 B,p_1A\sigma_1
B,p_2A\sigma_2B,p_3A\sigma_3B\}\label{gene}} with $A,B$ complex
$2\times 2$ matrices and $p_i\geq 0$, related to the $\{s_i\}$ by
the formula relating the eigenvalues of a Bell diagonal state to
its Lorentz singular values. The Kraus operators of maps with a
normal form of the second kind can be chosen to be of the form
{\small
\be{\{A_i\}=\{\sqrt{\frac{1+x}{2}}A\ba{cc}{1&0\\0&\frac{1}{\sqrt{3}}}B,\sqrt{\frac{1-x}{2}}A\ba{cc}{1&0\\0&-\frac{1}{\sqrt{3}}}B,
\sqrt{\frac{2}{3}}A\ba{cc}{0&1\\0&0}B\}\label{nge},}}again with
$A,B$ complex $2\times 2$ matrices. In the third case, the map is
trivial as it maps everything to the same point.
$\{s_i\},x,A,B,\{p_i\}$ can be calculated explicitly by
calculating the Lorentz singular value decomposition of the state
$\rho_\Phi$.
\end{theorem}
{\em Proof:} The proof is immediate given the Lorentz singular
value decomposition. The first case corresponds to a
diagonalizable $R$, and a diagonal $R$ corresponds to a
bistochastic channel. The second and third case correspond to
non-diagonalizable cases (note that there are $2$ normal forms in
the case of states that do not apply here as they cannot lead to
trace-preserving channels).\qed

This gives a nice classification of all the classes of TPCP-maps
on qubits: the generic class is the one that can be brought into
unital form by adding appropriate filtering transformations $A,B$,
i.e. the ellipsoid can be continuously deformed to an ellipsoid
whose center is the maximally mixed state. The non-generic class
however cannot be deformed in this way: it is easy to show that
the ellipsoid corresponding to the normal form touches the Bloch
sphere at one and only at one point; there is no filtering
operation that can change this property. We conclude that the
ellipsoids in the non-generic case are not (and cannot be made by
filtering operations) symmetric around the origin and that they
touch the Bloch sphere at exactly one point.

\begin{figure}
\epsfig{file=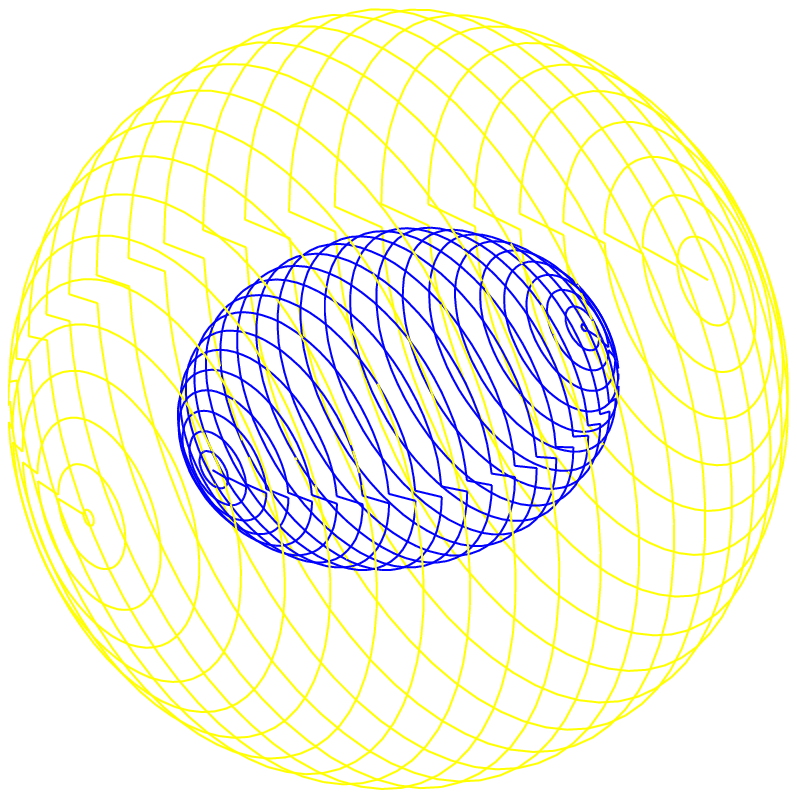,width=5cm} \epsfig{file=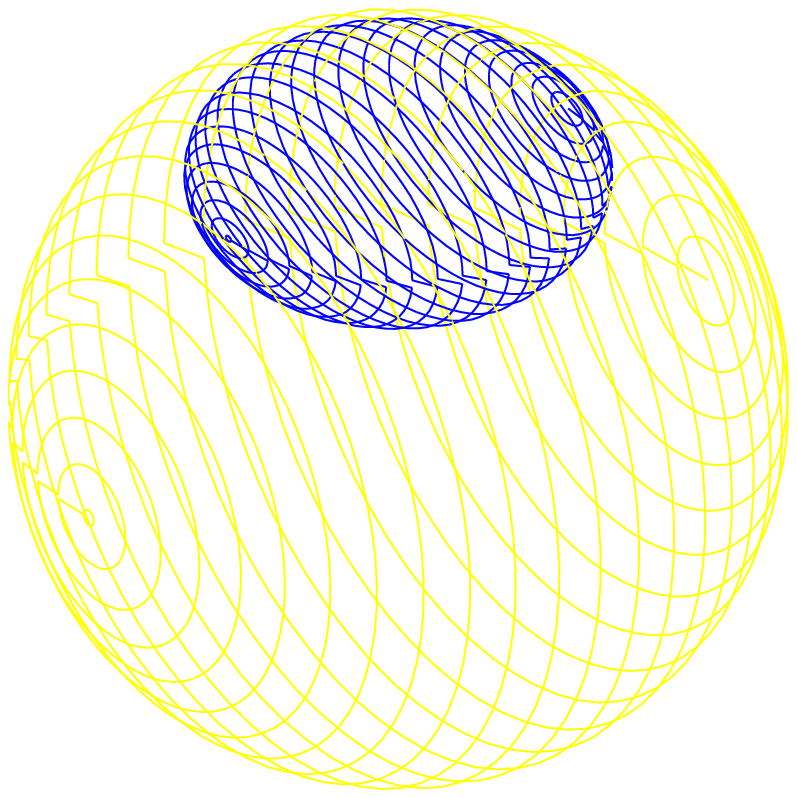,width=5cm}
\caption{\label{twokoe} The image of a channel in generic normal
form (left) or in non-generic normal form (right).}
\end{figure}

We depict both types of normal ellipsoids in figure \ref{twokoe}.
Note that this geometrical picture will be very useful in guessing
input states that maximize the classical capacity of the state
(see e.g. \cite{KR01}).

\subsection{Extremal maps for qubits}

In the case of a qubit channel $\Phi$, the dual state $\rho_\Phi$
is  a mixed state of two qubits. It is possible to obtain an
explicit parameterization of all extremal qubit maps (see also
Ruskai et al. \cite{RSW02} for  a different approach):
\begin{theorem}\label{th1}
The set of dual states $\rho_\Phi$ corresponding to extreme points
of the set of completely positive trace preserving maps $\Phi$ on
1 qubit is given by the union of all maximally entangled pure
states, and all rank 2 states $\rho$ for which ${\rm
Tr}_{2}(\rho_\Phi)$ is equal and ${\rm Tr}_{1}(\rho_\Phi)$ is not
equal to the identity. The Kraus operators corresponding to the
rank 1 extreme points are unitary, while the ones corresponding to
the rank 2 extreme points have a representation of the form:
\be{A_1=U\ba{cc}{s_0&0\\0&s_1}V^\dagger\hspace{1cm}
A_2=U\ba{cc}{0&\sqrt{1-s_1^2}\\\sqrt{1-s_0^2}&0}V^\dagger\label{extr2q}}
with $U,V$ unitary.
\end{theorem}
Proof: We have already proven  that extremal TPCP-maps have
maximal rank 2. Due to the duality between maps and states, it is
sufficient to consider rank 2 density operators of two qubits
$\rho_\Phi$ for which $Tr_2(\rho_\Phi)=I_2$. A real
parameterization of all 2-qubit density operators $\rho$ is given
by the real $4\times 4$ matrix $R$ with coefficients
\be{R_{ij}=Tr\left(\rho\sigma_i\otimes \sigma_j\right)} where
$0\leq i,j\leq 3$. An appropriate choice of local unitary bases
can always make the $R_{1:3,1:3}$ block diagonal, and the
trace-preserving condition translates into $R_{0,1:3}=0$.
Therefore $R$ is given by:
\[R=\ba{cccc}{1&0&0&0\\t_1&\lambda_1&0&0\\t_2&0&\lambda_2&0\\t_3&0&0&\lambda_3}.\]
The corresponding $\rho$ is given by
\[\rho=\frac{1}{4}\ba{cccc}{1+t_3+\lambda_3&       0& t_1-it_2&   \lambda_1-\lambda_2\\     0& 1+t_3-\lambda_3&
\lambda_1+\lambda_2& t_1-it_2\\  t_1+it_2&   \lambda_1+\lambda_2&
1-t_3-\lambda_3& 0\\\lambda_1-\lambda_2& t_1+it_2& 0&
1-t_3+\lambda_3},\] and the positivity of $\rho$ constrains the
allowed range of the 6 parameters. Let us now impose that the rank
of the corresponding $\rho$ is $2$. This implies that linear
combinations of $3\times 3$ minors of $\rho$ be zero, and after
some algebra one obtains the following conditions:
\begin{eqnarray*}
t_3(\lambda_3+\lambda_1\lambda_2)&=&0\\
t_2(\lambda_2+\lambda_1\lambda_3)&=&0\\
t_1(\lambda_1+\lambda_2\lambda_3)&=&0
\end{eqnarray*}
These equations, supplemented with the fact that diagonal elements
of a positive semidefinite matrix are always bigger than the
elements in the same column, lead to the conclusion that all $t_i$
but one have to  be equal to zero if $\rho$ is rank 2. Without
loss of generality, we can choose $t_1=t_2=0$ and parameterize
$\lambda_1=\cos(\alpha)$, $\lambda_2=\cos(\beta)$. We thus arrive
at the canonical form
\be{R=\ba{cccc}{1&0&0&0\\0&\cos(\alpha)&0&0\\0&0&\cos(\beta)&0\\\sin(\alpha)\sin(\beta)&0&0&-\cos(\alpha)\cos(\beta)}.\label{R2}}
Suppose that $\sin(\alpha)\sin(\beta)=0$ (this condition is
equivalent to $Tr_1(\rho_\Phi)=I/2$. Then the state corresponding
to this $R$ is Bell-diagonal and thus a convex sum of two
maximally entangled states, and therefore the map corresponding to
this state cannot be extremal. In the other case, an extremal rank
2 TPCP-map is obtained, which can easily be shown to yield the
given Kraus representation,  where
$s_0=\sqrt{1-\cos(\alpha+\beta)/2}$ and
$s_1=\sqrt{1-\cos(\alpha-\beta)}/2$.\qed

Note that the corresponding Theorem for bistochastic qubit
channels is not very useful, as extremal TPCP qubit channels are
always unitary. Theorem \ref{thepp} however is very interesting,
and indicates that there always exist pure states that remain pure
after the action of the extremal qubit channel: indeed, if the
basis vectors $\{|i\rangle\}$ are chosen according to the unitary
$V$ in (\ref{extr2q}), then it is easily checked that the states
$|\psi\rangle\simeq s_2\sqrt{1-s_2^2}|0\rangle\pm
s_1\sqrt{1-s_1^2}$ remain pure by the action of the extremal map.
Note that these two states are the only ones with this property,
and note also that they are not orthogonal to each other.

\subsection{Quantum capacity}

Let us now move on to the relation between 1-qubit quantum
channels and entanglement. We can now make use of the plethora of
results derived for mixed states of two qubits. Let us first
consider Theorem \ref{entbre} about entanglement breaking
channels. In the case of mixed states of two qubits, a state is
entangled iff it violates the reduction criterion $I\otimes
\rho_B-\rho\geq 0$. But in the case of the dual state $\rho_\Phi$,
it holds that $\rho_B=I/2$, and therefore it holds that a quantum
channel $\Phi$ can be used to distribute entanglement iff the
maximal eigenvalue of $\rho_\Phi$ exceeds $1/2$ \footnote{This was
first observed by Michael Horodecki}. In the light of Theorem
\ref{thme}, it follows that such a non-entanglement breaking
channel can always be used to distribute an entangled state with
fidelity larger than $1/2$, which implies on its turn that it can
be used to distill entanglement\cite{BDS96}.

Consider now an entanglement breaking channel, i.e. a channel for
which $\rho_\Phi$ is separable. In this case all the Kraus
operators can be chosen to be projectors. An explicit way of
calculating this Kraus representation exists. Indeed, in the
section about entanglement of formation of two qubits, a
constructive way of decomposing a separable mixed state of two
qubits as a convex combination of separable pure states was given.
It was furthermore proven that a separable state of rank 2 or 4
can always be written as a convex combination of 2 respectively 4
separable pure states, thus giving rise to 2 respectively 4 rank
one Kraus operators. Surprisingly, most separable rank 3 mixed
states of two qubits can only be written as a convex combination
of 4 separable pure states. This implies that a generic
entanglement breaking channel of rank 3 needs 4 Kraus operators if
these are to be chosen rank 1. Let us also mention that the set of
separable states is not of measure zero, implying that the set of
entanglement breaking channels is also not of measure zero.

The results of Wootters \cite{Woo98} can of course also be applied
to non-entanglement-breaking channels. A direct application of the
formalism of Wootters yields the following Theorem:

\begin{theorem}
Given a 1-qubit channel $\Phi$ and the state $\rho_\Phi$
associated to it. If $C$ is the concurrence of $\rho_\Phi$, then
the channel has a Kraus representation of the form:
\bea{\Phi(\rho)&=&\sum_i
p_i(U_i\tilde{C}V_i)\rho (U_i\tilde{C}V_i)^\dagger\\
\tilde{C}&=&\frac{1}{2}\ba{cc}{\sqrt{1+C}+\sqrt{1-C}&0\\0&\sqrt{1+C}-\sqrt{1-C}}}
where $U_i,V_i$ are unitary matrices.
\end{theorem}
Proof: The Theorem is a direct consequence of the fact that a
mixed state with concurrence $C$ can be written as a convex sum of
pure states all with concurrence equal to $C$.\qed

The geometrical meaning in the context of channels is the
following: each trace-preserving CP-map is a convex combination of
contractive maps in unique different directions, where each
contraction has the same magnitude.


Let us next address the question of calculating the quantum
capacity of the one-qubit channel. Clearly, Theorem \ref{thme}
tells us what states to send through the channel such as to
maximize the fidelity of the shared entangled states. In general,
the quantum capacity cannot be calculated as we even don't have a
way of calculating the entanglement of distillation of mixed
states of two qubits (which is a simpler problem).

In the case of unital channels of rank 2 however, the eigenvectors
of $\rho_\Phi$ are maximally entangled and the quantum capacity
can  be calculated explicitly:

\begin{theorem}
Consider a bistochastic qubit channel $\Phi$ of rank 2. Then its
quantum capacity is given by $C_Q=1-H(p)$, where $p$ is the
maximal eigenvalue of $\rho_\Phi$  and
$H(p)=-p\log_2(p)-(1-p)\log_2(1-p)$.
\end{theorem}
Proof: A unital qubit channel exhibits the nice property that no
loss whatever occurs by sending a maximally entangled state
through the channel: it can easily be shown (see Bennett et
al.\cite{BDS96}) that sending a quantum system through the channel
is equivalent to using the standard teleportation channel induced
by the (non-maximally entangled state) $\rho_\Phi$. Because we can
use the state $\rho_\Phi$, obtained by sending a Bell state
through the channel, to perfectly simulate the channel, this is
clearly the optimal thing to do, and the quantum capacity of the
channel is therefore equal to the distillable entanglement of
$\rho_\Phi$. Now Rains \cite{Rai99} has proven that the
distillable entanglement of a Bell diagonal state of rank 2 is
given by $E_{dist}(\rho)=1-S(\rho)$, which ends the proof of the
Theorem.\qed

More general, the quantum capacity of bistochastic qubit channel
is always equal to the entanglement of distillation of the
corresponding dual states (due to the  arguments in the previous
proof).

As a last remark, we observe that the channels of the non-generic
kind that touch the Bloch sphere at exactly one point are never
entanglement-breaking: this follows from the fact that the
concurrence of $\rho_\Phi$ always exceeds $0$ in that case.

\subsection{Classical capacity}

Far more progress has been made concerning the classical capacity
of quantum channels: it is known that the classical capacity using
product inputs is given by the Holevo-$\chi$ quantity. Here the
geometrical picture derived in  section 6.4 can sharpen our
intuition. Consider for example the case of a unital channel. It
is immediately clear that Holevo-$\chi$  will be maximized by
choosing a mixture of two states that lie on the opposite side of
the major axis of the ellipsoid. This implies that the optimal
input states are orthogonal. King and Ruskai \cite{KR01,Kin01}
even proved that entangled inputs cannot help in the case of
unital channels, and we conclude that the classical capacity of
the unital channels is completely understood.

Consider however  a non-unital channel of the generic kind. As
proven before, this channel can be interpreted as the succession
of a filter, a unital channel, and another filter.  The critical
source of noise or decoherence and irreversibility in a channel is
the mixing, and the previous analysis tells us that this mixing
can always be interpreted to happen in a unital way, whereas the
in- and output of the unital channel is reversibly but
non-orthogonally filtered. It follows that orthogonal inputs will
not appear orthogonally in the unital channel, and typically
orthogonal inputs will not achieve capacity. This strange fact was
indeed discovered by Fuchs \cite{Fuc97}, and it appears to be
generic for non-unital channels.

Let us now have a look at the non-generic family of channels,
whose ellipsoids touch the Bloch sphere at exactly one point. It
happens that the so-called stretched channel belongs to this
family, and this channel has the property that its (product)
capacity is only achieved for an input ensemble with three
states\cite{KNR02}. This is surprising but not too surprising
given the geometrical picture, as one of the input states
corresponds to the pure output state, while the other two ones are
chosen to lie symmetric around the axis connecting the maximally
entangled state with the pure output state. Note however that most
of the non-generic states achieve capacity with 2 input states.

Let us now move to calculate the classical capacity of the
extremal qubit channels.  In the case of extremal qubit channels,
it is possible to reduce the problem of calculating the classical
(Holevo) capacity to an optimization problem over the ensemble
average. The problem to be solved is as follows: find the optimal
ensemble $\{\rho_i,p_i\}$ such that
\[S(\sum_i p_i\Phi(\rho_i))-\sum_ip_iS(\Phi(\rho_i))\]
is maximized. We assume that $\Phi$ is rank 2 and therefore has a
Kraus representation of the form (\ref{extr2q}). It is clear that
only pure states $\{\rho_i\}$ have to be considered. It is easily
seen that in the case of qubits, the entropy of a state is a
convex monotonously increasing function of the determinant of the
density operator: $S(\rho)=H(1/2(1-\sqrt{1-4\det(\rho)^2}))$ with
$H(p)=p\log(p)+(1-p)\log(1-p)$ the Shannon entropy function.
Inspired by the analysis of 2-qubit channels by Uhlmann in terms
of anti-linear operators \cite{Uhl01}, we make the following
observation: \be{\det\left(A_1|\psi\rangle\langle\psi|A_1^\dagger
+
A_2|\psi\rangle\langle\psi|A_2^\dagger\right)=|\psi^T(A_1^T\sigma_y
A_2-A_2^T\sigma_yA_1)\psi|.} Here $\psi$ is the vector notation
(in the computational basis) of $|\psi\rangle$, and $\sigma_y$ is
a Pauli matrix. Suppose now that we add an additional constraint
to the problem, namely that the  ensemble average $\rho$ is given.
Taking a square root $X$ of $\rho=XX^\dagger$, all possible pure
state decompositions can be written as $X'=XU$ with $U$ an
arbitrary isometry (note that the columns of $XU$ represent all
unnormalized pure states in the decomposition). With this
additional constraint, the problem can be solved exactly as we
solved the entanglement of formation problem. A constructive way
of obtaining the optimal decomposition of $\rho$ is as follows:
take a square root $X$ of $\rho$, and calculate the singular value
decomposition of the symmetric matrix $X^T(A_1^T\sigma_y
A_2-A_2^T\sigma_yA_1)X=V\Sigma V^T$. Call $C=\sigma_1-\sigma_2$
the concurrence with $\{\sigma_i\}$ the singular values of the
above symmetric matrix. Then the optimal decomposition is obtained
by choosing $U=V^*O$ with $O$ the real orthogonal matrix that is
chosen such that the diagonal entries of the matrix $R=O^T({\rm
Diag}[\sigma_1,- \sigma_2]-C\rho)O)$ vanish. For given ensemble
average $\rho$, the classical capacity is therefore given by the
following formula: $S(\Phi(\rho))-f(C)$ (see also Uhlmann \cite{Uhl01}).\\

To derive an explicit formula for the classical capacity of the
extremal channels, we still have to do an optimization over all
possible ensemble averages $\rho$. Note that the previous analysis
already learned us that the capacity will always be reached with
an ensemble of two input states. Both the terms $\Phi(\rho)$ and
$C$ can easily be extremized separately, but unfortunately even if
the eigenvalues of $\rho$ are fixed, the optimal eigenvectors for
maximizing $S(\rho)$ and minimizing $C$ are not compatible.
However, the capacity can easily be calculated numerically, as it
just an optimization problem over three real parameters.

On the other hand, we have seen that the definition of the
classical capacity had a direct counterpart in giving an appealing
definition for the number of classical correlations present in a
(mixed) bipartite state $C_{cl}$ (see \ref{Ecl}). The techniques
used in the foregoing paragraph are perfectly adequate to give an
exact expression of this quantity if the shared quantum state is a
rank 2 bipartite state $\rho$ of qubits. Indeed, a mixed bipartite
state of two qubits can just be seen as a more general kind of
quantum channel.

\section{Conclusion}
We have shown that the natural description of quantum channels or
positive linear maps is given by a dual quantum state associated
to the map. This dual state is defined over a Hilbert space that
is naturally endowed with a tensor product structure of the in-
and output of the channel. We showed that the techniques developed
in the context of entanglement are of direct use in describing
positive maps. We derived a characterization of the extreme points
of the convex set of trace-preserving completely positive maps,
and gave some generalizations. We discussed some new results about
the classical and quantum capacity of a quantum channel, and in
the case of one-qubit channels we showed how to exploit the
duality between qubit channels and mixed states of two qubits to
obtain useful parameterizations.

\bibliographystyle{unsrt}

\end{document}